\documentclass[12pt,reqno]{article}
\usepackage{fullpage}
\usepackage[usenames]{color}
\usepackage{amscd}
\usepackage[colorlinks,linkcolor=webgreen,filecolor=webbrown,citecolor=webgreen]{hyperref}\usepackage{breakurl}
\definecolor{webgreen}{rgb}{0,.5,0}
\definecolor{webbrown}{rgb}{.6,0,0}
\usepackage{amssymb,amsmath,amsthm}
\usepackage{epigraph}
\usepackage{pstricks,multido}
\usepackage{graphicx}
\newtheorem*{thm}{Theorem}
\usepackage{xspace}
\let\Sect=\S
\newcommand{\N}{\textsf{N}\xspace}
\renewcommand{\S}{\textsf{S}\xspace}
\newcommand{\E}{\textsf{E}\xspace}
\newcommand{\W}{\textsf{W}\xspace}
\newcommand{\U}{\textsf{U}\xspace}
\newcommand{\D}{\textsf{D}\xspace}
\newcommand{\NN}{\textsf{NN}\xspace}
\newcommand{\NS}{\textsf{NS}\xspace}
\newcommand{\SN}{\textsf{SN}\xspace}
\newcommand{\NE}{\textsf{NE}\xspace}
\newcommand{\SE}{\textsf{SE}\xspace}
\renewcommand{\SS}{\textsf{SS}\xspace}
\newcommand{\head}[1]{\multicolumn{1}{|c|}{\bf #1}}

\def\div#1#2{\lfloor#1/#2\rfloor}
\newcommand{\seqnum}[1]{\href{http://oeis.org/#1}{\underline{#1}}}
\title{Touchard's Drunkard}
\author{Nachum Dershowitz\thanks{Corresponding author. This 
research benefitted from a fellowship at the Paris Institute for Advanced Studies (France), with the financial support of the French state, managed by the French National Research Agency's ``Investissements d'avenir'' program (ANR-11-LABX-0027-01 Labex RFIEA+).}\\
School of Computer Science\\Tel Aviv University\\Ramat Aviv, Israel\\
\href{mailto:nachum@cs.tau.ac.il}{\tt nachum@cs.tau.ac.il}}
\begin{document}
\maketitle

\epigraph{\small You're a baby and as stupid as a Frenchman. You persist in thinking that it's the same as it was at Touchard's, and that I'm as stupid as at Touchard's\dots. But I'm not so silly as I was at Touchard's\dots. I was drunk yesterday, but not from wine, but because I was excited.}{---Fyodor Dostoyevsky, \textit{The Raw Youth} (1875)}

\begin{abstract}
Based on Touchard's identity, a simple derivation is given for the enumeration of the  \N/\S/\E/\W walks that remain on the north side of the origin.
\end{abstract}

\section{Introduction: Drunken walks}

\begin{figure}
\centering
\includegraphics[width=0.85\linewidth]{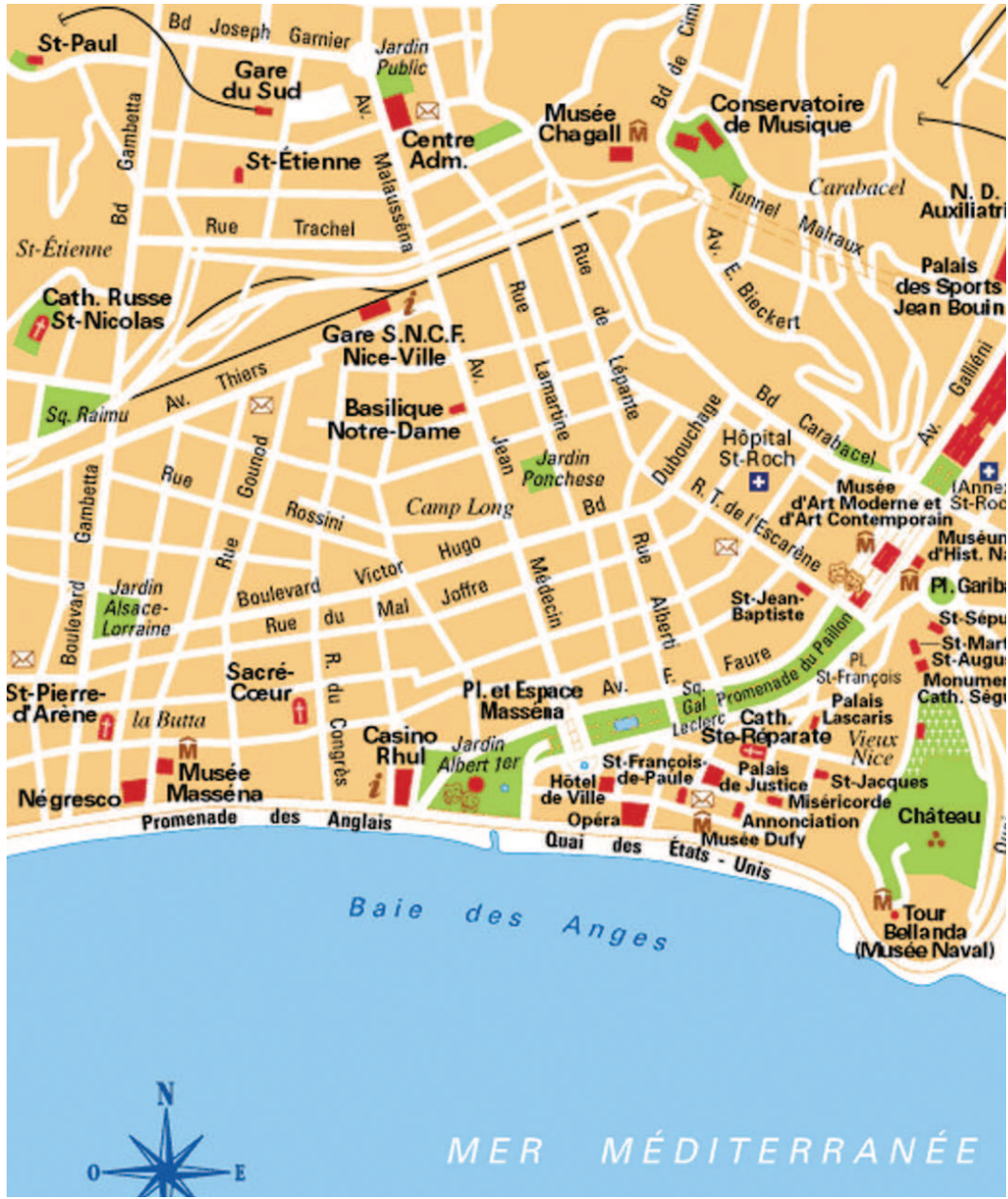}
\caption{Center of Nice, France, with the promenade at its southern end.}\label{fig:Nice}
\end{figure}

An inebriated person in Nice (see Figure~\ref{fig:Nice}) takes a walk, each step in one of the four cardinal directions, north (\N), south (\S), east (\E), and west (\W).  
We are interested in those walks beginning at the center of the Promenade des Anglais (at the southern end of town\footnote{And site of recent carnage.}) and
ending anywhere on the promenade---all the while remaining on land (in other words, not venturing south of the promenade).
In how many possible ways can such walks meander? 
 Let
\begin{align*}
D_n &= \left\{ ~
\parbox{11cm}{the number of \emph{Touchard walks}, consisting of a sequence of $n$ steps, each of which is one of \N/\S/\E/\W, such that at each point along the way the number of \N-steps that have been taken is never less than the number of \S-steps,
and, furthermore, in the end they are equal (with no restrictions on the distribution of \E- or \W-steps).}
\right.
\end{align*}
These are walks that remain in the half-plane and return to the boundary.
Table~\ref{tab}
lists valid and invalid walks of length $n=4$.
For an example of a longer walk, see Figure~\ref{fig:walk}.

\begin{thm}
The number of Touchard walks with a total of $n$ steps is
\begin{align*}
D_n &=C_{n+1} ,
\end{align*} 
where $C_i$ is the $i$th Catalan number, $\frac{1}{i+1}{2i\choose i}$ {\rm(sequence \seqnum{A000108} in Sloane's \textit{Encyclopedia of Integer Sequences}~\cite{EIS})}.
\end{thm}

\begin{table}
\centering
\begin{tabular}{|llll|llll|}
\multicolumn{4}{|c|}{\bf Valid} & \multicolumn{4}{c|}{\bf Invalid} \\\hline
\N& \N& \S& \S & \N& \S& \underline \S&  \N\\
\N& \S& \N& \S &  \underline \S&  \N& \N& \S\\\hline
\N& \S& \E& \E & \underline \S&  \N& \E & \E\\
\N& \S& \E& \W & \underline \S&  \N& \E & \W\\
\N& \S& \W& \E & \underline\S& \N & \W& \E \\
\N& \S& \W& \W & \underline\S& \N & \W& \W \\
\N& \E& \S& \E & \underline\S& \underline\E& \N & \E \\
\N& \E& \S& \W & \underline\S&\underline \E& \N & \W \\
\N& \W& \S& \E & \underline\S&\underline \W& \N & \E \\
\N& \W& \S & \W& \underline\S&\underline \W& \N & \W\\
\N& \E& \E& \S & \underline\S&\underline \E& \underline\E& \N \\
\N& \E& \W& \S & \underline\S&\underline \E& \underline\W& \N \\
\N& \W& \E& \S & \underline\S&\underline \W& \underline\E& \N \\
\N& \W& \W& \S & \underline\S&\underline \W& \underline\W& \N \\\hline
\end{tabular}
\qquad
\begin{tabular}{|llll|llll|}
\multicolumn{4}{|c|}{\bf Valid} & \multicolumn{4}{c|}{\bf Invalid} \\\hline
&&&& \underline \S&  \N& \underline\S & \N\\
&&&& \underline \S & \underline \S & \underline \N & \N\\\hline
\E& \N& \S& \E & \E& \underline\S& \N & \E \\
\E& \N& \S& \W & \E& \underline\S& \N & \W \\
\W& \N& \S& \E & \W& \underline\S& \N & \E \\
\W& \N& \S& \W & \W& \underline\S& \N & \W \\
\E& \N& \E& \S & \E& \underline\S&\underline \E& \N \\
\E& \N& \W& \S & \E& \underline\S&\underline \W& \N \\
\W& \N& \E& \S & \W& \underline\S&\underline \E& \N \\
\W& \N& \W& \S & \W& \underline\S&\underline \W& \N \\
\E& \E& \N& \S & \E& \E& \underline\S& \N \\
\E& \W& \N& \S & \E& \W& \underline\S& \N \\
\W& \E& \N& \S & \W& \E& \underline\S& \N \\
\W& \W&  \N&\S & \W& \W&  \underline\S&\N \\\hline
\end{tabular}
\caption{All Touchard walks of length 4 with equal quantities of \N-steps and \S-steps, 26 valid and 28 not, besides 16 valid \E/\W walks with no \N/\S-steps at all. 
The illegal steps in the Mediterranean 
are \underline{underlined}.
Walks with unequal numbers of \N- and \S-steps are always invalid.}
\label{tab}
\end{table}

\begin{figure}
\centering
\begin{pspicture}(0,4.5)
\psgrid[subgriddiv=1,griddots=10](0,0)(-5,0)(5,4)
\psline[doubleline=true,linestyle=dashed]{-}(-5,0)(5,0)
\psset{linecolor=blue,linewidth=2pt}
\psline{->}(0,0)(0,1)
\psline{->}(0,1)(1,1)
\psline[linearc=.05]{->}(1,1)(1.1,1.05)(1,1.1)(0,1.1)
\psline{->}(0,1.1)(-1,1.1)
\psline{->}(-1,1.1)(-1,2)
\psline{->}(-1,2)(-1,3)
\psline{->}(-1,3)(0,3)
\psline{->}(0,3)(1,3)
\psline{->}(1,3)(1,2)
\psline{->}(1,2)(1.9,2)
\psline{->}(1.9,2)(1.9,3)
\psline{->}(1.9,3)(1.9,4)
\psline[linearc=.05]{->}(1.9,4)(1.95,4.1)(2,4)(2,3)
\psline{->}(2,3)(2,2)
\psline{->}(2,2)(2,1)
\psline{->}(2,1)(2,0)
\psline{->}(2,0)(3,0)
\psline{->}(3,0)(4,0)
\end{pspicture}
\medskip
\caption{A valid walk, \N\E\W\W\N\N\E\E\S\E\N\N\S\S\S\S\E\E, consisting of 18 steps, 5 \N, 5 \S, 6 \E, and 2 \W.}\label{fig:walk}
\end{figure}
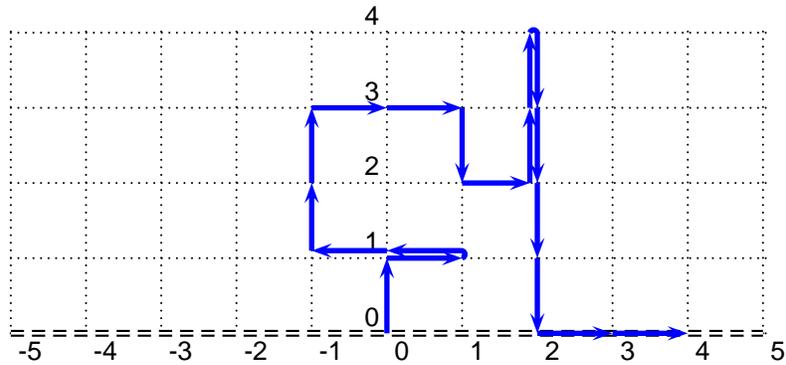

\begin{figure}
\centering
\begin{pspicture}(10,4.5)
\psgrid[subgriddiv=1,griddots=10](0,0)(0,0)(10,4)
\psset{linecolor=red}
\psline[linestyle=dashed]{->}(0,0)(1,1)
\psline[linestyle=dashed]{->}(1,1)(2,2)
\psline[linestyle=dashed]{->}(2,2)(3,3)
\psline[linestyle=dashed]{->}(3,3)(4,2)
\psline[linestyle=dashed]{->}(4,2)(5,3)
\psline[linestyle=dashed]{->}(5,3)(6,4)
\psline[linestyle=dashed]{->}(6,4)(7,3)
\psline[linestyle=dashed]{->}(7,3)(8,2)
\psline[linestyle=dashed]{->}(8,2)(9,1)
\psline[linestyle=dashed]{->}(9,1)(10,0)
\psset{linecolor=blue,linewidth=2pt}
\psline{->}(0,0)(0,1)
\psline{->}(1,1)(1,2)
\psline{->}(2,2)(2,3)
\psline{->}(3,3)(3,2)
\psline{->}(4,2)(4,3)
\psline{->}(5,3)(5,4)
\psline{->}(6,4)(6,3)
\psline{->}(7,3)(7,2)
\psline{->}(8,2)(8,1)
\psline{->}(9,1)(9,0)
\end{pspicture}
\medskip
\caption{The ten \N/\S-steps of Figure~\ref{fig:walk}, stretched out on a timeline: \N\N\N\S\N\N\S\S\S\S.
Connecting the tails of the steps yields a Dyck path of \NE/\SE steps.}\label{fig:NS}
\end{figure}
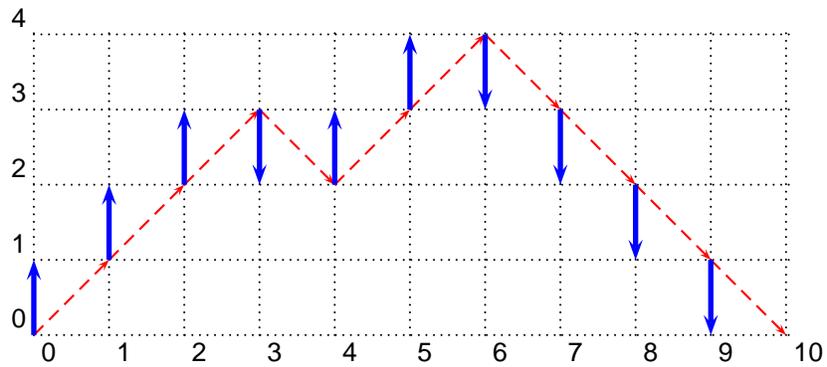

\begin{figure}
\centering
\begin{pspicture}(0,8.5)
\psgrid[subgriddiv=1,griddots=10](0,0)(-5,0)(5,8)
\psset{linecolor=red}
\psline[linestyle=dashed]{->}(0,0)(1,1)
\psline[linestyle=dashed]{->}(1,1)(0,2)
\psline[linestyle=dashed]{->}(0,2)(-1,3)
\psline[linestyle=dashed]{->}(-1,3)(0,4)
\psline[linestyle=dashed]{->}(0,4)(1,5)
\psline[linestyle=dashed]{->}(1,5)(2,6)
\psline[linestyle=dashed]{->}(2,6)(3,7)
\psline[linestyle=dashed]{->}(3,7)(4,8)
\psset{linecolor=blue,linewidth=2pt}
\psline{->}(0,0)(1,0)
\psline{->}(1,1)(0,1)
\psline{->}(0,2)(-1,2)
\psline{->}(-1,3)(0,3)
\psline{->}(0,4)(1,4)
\psline{->}(1,5)(2,5)
\psline{->}(2,6)(3,6)
\psline{->}(3,7)(4,7)
\end{pspicture}
\medskip
\caption{The eight \E/\W-steps of Figure~\ref{fig:walk}: \E\W\W\E\E\E\E\E.}\label{fig:EW}
\end{figure}

There are, for instance, $C_5=42$ valid 4-step walks,  listed in Table~\ref{tab}.

Guy~\cite{Guy}  points out that this equality ``is not well known!\@ \dots\@ nor can we immediately see any correspondence between [Touchard] walks and any of the manifestations [of Catalan objects].''
We aim to fill this lacuna.

For a history of Catalan enumerations, see \cite[Appendix B]{Stanley}.
 
\section{Enumeration: Touchard's identity}

Touchard's \cite{Touchard} identity (see, for example, \cite[p.~319]{Koshy}) states that
\begin{align*}
C_{n+1} &= \sum_i C_i 2^{n-2i} {n\choose 2i} .
\end{align*}
For a nice proof of this, see \cite{Shapiro}.

Considering the above theorem and this identity, we can understand the drunken
walks in Table~\ref{tab}, for $n=4$, 
as comprising $\frac{1}{1}{0\choose 0} 2^{4} {4\choose 0}=16$ valid walks with no ($i=0$) north-south steps,  $\frac{1}{2}{2\choose 1} 2^{2} {4\choose 2}=24$ walks containing one ($i=1$) north-step followed at some point by one south-step,
and
$\frac{1}{3}{4\choose 2} 2^{0} {4\choose 4}=2$ walks with two ($i=2$) north-steps and two matching south-steps.

With Touchard's identity, the proof of the theorem is  immediate: 
\begin{enumerate}
\item\label{s1} Suppose there are $i$ \N-steps and $i$ \S-steps, for some $i$ in the range $[0\mathbin{..} n/2]$, leaving $n-2i$ steps of type \E or \W.
\item\label{s2} The factor $C_i$ counts the patterns consisting of $i$ \N-steps and an equal number of \S-steps, starting and ending on the promenade, and never venturing further south (see Figure~\ref{fig:NS}).   This is one of the many well-known instances of Catalan enumerations, and is a special case of the famous ``ballot problem,'' stated and solved by Whitworth back in  1878
\cite{Whitworth}.
\item\label{s3} The factor  $2^{n-2i}$ counts the patterns of the remaining unconstrained \E/\W-steps  (see Figure~\ref{fig:EW}).
\item\label{s4} The factor ${n\choose 2i}$ is the number of ways of interspersing  $2i$ \N/\S-steps among  $n-2i$ \E/\W-steps.
\end{enumerate}

\section{Bijection: Dyck paths}

Walks whose steps  are only north or south and stay on land (as in step~\ref{s2} in the above proof) correspond to the well-known Dyck (monotonic lattice) paths~\cite[pp.\@ 151--153]{Dyck,Koshy}, which spread out the steps over a timeline that runs from west to east.
Dyck paths are usually depicted as consisting of equal numbers of \NE- and \SE-steps,
staying the whole time north of the origin. They are counted by the Catalan numbers.
For an example, see the dashed line in Figure~\ref{fig:NS}.
Alternatively, such paths may be viewed as consisting of \N- and \E-steps, never going below the $y=x$ diagonal;
see~\cite[pp.\@1--4]{lattice}.

Alternatively, Dyck paths may be viewed as consisting of \N/\S-steps
meeting the requirements that (a) the number of  south steps---throughout the walk---never exceeds the number of north ones
and (b) that---at the end---they  be equal.
The enumeration $D_n$ counts walks in any of the four directions (\N/\S/\E/\W)
abiding by the identical constraints.

\begin{figure}
\newcommand{\x}{\hspace{7pt}}
\newcommand{\y}{\hspace{6pt}}
\centering
\N\x\E\x\W\y\W\y\N\x\N\x\E\x\E\x\S\x\E\x\N\x\N\x\S\x\S\x\S\x\S\x\E\x\E\\
$\Leftrightarrow$\\
\N\NN\NS\SN\SN\NN\NN\NS\NS\SS\NS\NN\NN\SS\SS\SS\SS\NS\NS\S

\begin{pspicture}(16,4.3)
\psset{unit=4.15mm}
\psgrid[subgriddiv=1,griddots=10,gridlabels=7pt](0,0)(0,0)(38,9)
\psset{linecolor=blue}
\multido{\nx=0+1,\ny=0+1,\nxx=1+1,\nyy=1+1}{4}{\psline{->}(\nx,\ny)(\nxx,\nyy)}
\multido{\nx=4+1,\ny=4+-1,\nxx=5+1,\nyy=3+-1}{2}{\psline{->}(\nx,\ny)(\nxx,\nyy)}
\multido{\nx=6+1,\ny=2+1,\nxx=7+1,\nyy=3+1}{1}{\psline{->}(\nx,\ny)(\nxx,\nyy)}
\multido{\nx=7+1,\ny=3+-1,\nxx=8+1,\nyy=2+-1}{1}{\psline{->}(\nx,\ny)(\nxx,\nyy)}
\multido{\nx=8+1,\ny=2+1,\nxx=9+1,\nyy=3+1}{6}{\psline{->}(\nx,\ny)(\nxx,\nyy)}
\multido{\nx=14+1,\ny=8+-1,\nxx=15+1,\nyy=7+-1}{1}{\psline{->}(\nx,\ny)(\nxx,\nyy)}
\multido{\nx=15+1,\ny=7+1,\nxx=16+1,\nyy=8+1}{1}{\psline{->}(\nx,\ny)(\nxx,\nyy)}
\multido{\nx=16+1,\ny=8+-1,\nxx=17+1,\nyy=7+-1}{3}{\psline{->}(\nx,\ny)(\nxx,\nyy)}
\multido{\nx=19+1,\ny=5+1,\nxx=20+1,\nyy=6+1}{1}{\psline{->}(\nx,\ny)(\nxx,\nyy)}
\multido{\nx=20+1,\ny=6+-1,\nxx=21+1,\nyy=5+-1}{1}{\psline{->}(\nx,\ny)(\nxx,\nyy)}
\multido{\nx=21+1,\ny=5+1,\nxx=22+1,\nyy=6+1}{4}{\psline{->}(\nx,\ny)(\nxx,\nyy)}
\multido{\nx=25+1,\ny=9+-1,\nxx=26+1,\nyy=8+-1}{8}{\psline{->}(\nx,\ny)(\nxx,\nyy)}
\multido{\nx=33+1,\ny=1+1,\nxx=34+1,\nyy=2+1}{1}{\psline{->}(\nx,\ny)(\nxx,\nyy)}
\multido{\nx=34+1,\ny=2+-1,\nxx=35+1,\nyy=1+-1}{1}{\psline{->}(\nx,\ny)(\nxx,\nyy)}
\multido{\nx=35+1,\ny=1+1,\nxx=36+1,\nyy=2+1}{1}{\psline{->}(\nx,\ny)(\nxx,\nyy)}
\multido{\nx=36+1,\ny=2+-1,\nxx=37+1,\nyy=1+-1}{2}{\psline{->}(\nx,\ny)(\nxx,\nyy)}
\end{pspicture}
\caption{The Dyck path corresponding to the Touchard walk of Figure~\ref{fig:walk}.}\label{fig:Dyck}
\end{figure}
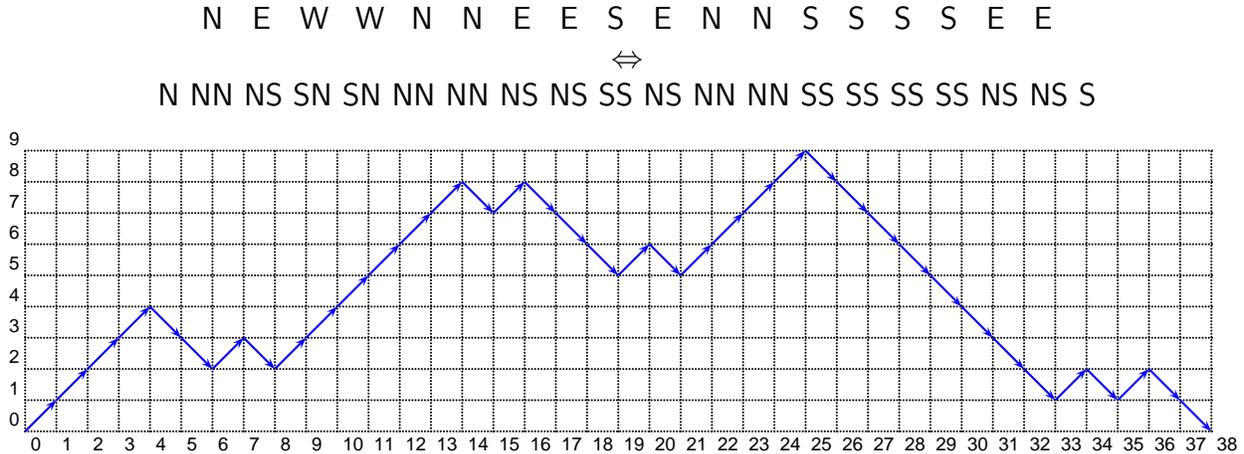

To relate the two kinds of  walks,
consider a Dyck path of length $2n+2$,
of which there are $C_{n+1}$.  It must start with \N and end with \S.  Forget those two steps.  Then start from the beginning and replace as follows: $\NN\mapsto\N$, $\SS\mapsto\S$, $\NS\mapsto\E$, $\SN\mapsto\W$. 
The result is a Touchard walk of length $n$.
The reverse direction of this bijection is straightforward.
See Figure~\ref{fig:Dyck}.

This construction is similar to one used in~\cite{Breck,GKS}.
Touchard walks are also easily seen to be in bijection with two-colored Motzkin paths \cite{two}
(the two colors being \E and \W).  
These in turn are in bijection with 
ballot sequences~\cite{Fell} or Dyck paths~\cite{Dyck}
in a manner similar to the above;
see \cite[item 40]{Stanley}.

\section{Extension: More dimensions}

Walks can be entertained in more dimensions with varying degrees of restriction.
\begin{enumerate}\renewcommand{\theenumi}{\textit{\alph{enumi}}}

\item\label{a} For each dimension with its two opposing directions (like \N and \S above) that must stay to one side 
of the origin and return to  zero (the promenade in our example) at the end, there is a Catalan factor 
\begin{align*}
C_i &= \frac{1}{i+1}{2i\choose i},
\end{align*}
accounting for $2i$ steps, $i$ in each direction (\seqnum{A000108}).
This component represents ballot sequences or Dyck paths.

\item\label{b} For each dimension that must return to zero at the end
(but need not stay on one side
of the origin), there is a central binomial factor
\begin{align*}
A_{2i} &= {2i \choose i}
\end{align*}
for $i$ steps in each of the two directions, in any order (\seqnum{A000984}).
These are the linear ``drunken walks'' studied by Poly\'a~\cite{Polya}, also called ``grand-Dyck'' paths~\cite{GD}.

\item\label{c} For each dimension that must stay on one side
of the origin (but need not return to zero at the end), there is a central binomial factor
\begin{align*}
A_{j} &= {j \choose \div{j}{2}}
\end{align*}
for a total of $j$ steps (\seqnum{A001405}).
This component represents ballot sequences with an uneven number of votes for the two candidates or prefixes of Dyck paths.

\item\label{d} For any remaining $r$ unrestricted directions (like \E and \W above), 
there is an exponential factor 
\begin{align*}
r^{n-m}
\end{align*}
covering the $n-m$ steps that are not yet accounted for, where $m=2i_1+2i_2+\cdots+j_1+\cdots$, the $i_k$ for each case (\ref{a}) or  (\ref{b}) and the $j_k$ for cases (\ref{c}).
When all directions are accounted for by cases (\ref{a}--\ref{c}) and $r=0$,  we must have $m=n$ and this factor is $1$.

\item[$\bullet$] To fix which of the $n$ steps belong to which category, there is a multinomial choice
\begin{align*}
{n \choose {2i_1,\dots,j_1,\dots,n-m}}.
\end{align*}
The steps in dimensions adhering to cases (\ref{a}) and  (\ref{b}) have an even number $2i_k$ of steps;
cases (\ref{c}) and  (\ref{d}) can have an odd number $j_k$ of steps.

\item[$\bullet$]  All the factors are summed for all possible values of the indices:
\begin{align*}
\sum_{i_1,i_2,\dots,j_1,j_2,\dots} & r^{n-m} C_{i_1} C_{i_2} \cdots A_{2i_\ell} \cdots  A_{j_1} \cdots {n \choose {2i_1,2i_2,\dots,j_1,\dots,n-m}},
\end{align*}
where $m=2i_1+2i_2+\cdots+j_1+j_2+\cdots$.
If $r=0$, however, the sum is
\begin{align*}
\sum_{m=n} & C_{i_1} C_{i_2} \cdots  A_{2i_\ell} \cdots  A_{j_1} \cdots {n \choose {2i_1,2i_2,\dots,j_1,\dots}}.
\end{align*}
\end{enumerate}

We can indicate the type of walk by a multiset of letters for the relevant cases.
Each dimension contributes a letter \textit{a}--\textit{e}, where \textit{e} is short for \textit{dd}, meaning that there are no restrictions on steps in that dimension,
whereas \textit{d} means that the dimension is one-way only.
Our drunkard's walk, then, is of type \textit{ae}, being confined to the northern half of the plane
but unrestricted longitudinally.

One-dimensional paths of types \textit{a, b, c, ee} are classified in \cite{BF} as \textit{excursions, bridges, meanders, and  walks}, respectively,
based on terminology of the theory of Brownian motion.

Whenever there are only dimensions of types \textit{a} and \textit{b}, the number of walks is 0 for an odd number $n$ of steps.

Motzkin paths~\cite[pp.\@ 300--301]{course} are like Dyck paths but allow arbitrary horizontal \E-steps in addition to \NE and \SE.
They are equivalent to walks of type \textit{ad} and are enumerated by the Motzkin numbers (\seqnum{A001006}),
\begin{align*}
M_n &= \sum_{i}C_i  {n\choose 2i} .
\end{align*}

Were we to insist that our drunkard return to the origin at the end of an evening of 
wanderings, then those would be walks of type \textit{ab},
which are counted by
\begin{align*}
\sum_{i} C_{\frac n 2 -i} {n\choose 2i}  {2i\choose i} &= C_{\frac n 2} {n+1 \choose n/2} 
\end{align*}
for even $n$~\cite{ab}.
This is \seqnum{A000891}$(n/2)$.
When $n$ is odd, there is---of course---no way home.
(Nagy~\cite{Nagy} finds related formul{\ae} for the case when an \N/\S-walk crosses the
abscissa an even number of times going south.)

The simplification of the above sum for walks of type \textit{ab}, as well as  the next three, 
may be seen as the result of a few applications of binomial cancellation (the ``subset of subsets equation'')~\cite[eq.~1.2.6(20)]{Knuth} followed by Vandermonde's convolution~\cite[eq.~3.1]{Gould}:
\begin{align*}
\sum_{i} \frac{1}{\frac{n}{2} -i+1}{n-2i\choose n/2 -i} {n\choose 2i}  {2i\choose i} 
&= \sum_{i} \frac{1}{\frac{n}{2} -i+1}{n-2i\choose n/2 -i} {n\choose i}  {n-i\choose i} \\
&= \sum_{i} \frac{1}{\frac{n}{2} -i+1}{n\choose i } {n-i \choose n/2-i}  {n/2 \choose i}  \\
&= \sum_{i} \frac{1}{\frac{n}{2} -i+1}{n\choose n/2 } {n/2 \choose i}  {n/2 \choose i}  \\
&= \frac{1}{\frac{n}{2}+1} {n\choose n/2 } \sum_{i}  {n/2 \choose i}  {n/2  + 1\choose n/2-i}  \\
&= \frac{1}{\frac{n}{2}+1} {n\choose n/2} {n+1 \choose n/2} .
\end{align*}

\begin{table}
\centering\renewcommand{\arraystretch}{1.4}
\begin{tabular}{l ||c|c|c|c|c|}
& \textit{a} & \textit{b} & \textit{c} & \textit{d} & \textit{e}  \\\hline\hline
\textit{a} 
&
\seqnum{A005568}* & 
\seqnum{A000891}* & 
\seqnum{A001700}  & 
 \seqnum{A001006} & 
\seqnum{A000108} 
\\\hline
\textit{b} 
&&
 \seqnum{A002894}* & 
 \seqnum{A018224} & 
\seqnum{A002426} & 
\seqnum{A000984}  
\\\hline
\textit{c} 
&&&
\seqnum{A005566} & 
\seqnum{A005773} & 
\seqnum{A001700} 
\\\hline
\textit{d} 
&&&&
\seqnum{A000079} & 
\seqnum{A000244} 
\\\hline
\textit{e} 
&&&&&
\seqnum{A000302} 
\\\hline
\end{tabular}
\caption{Two-dimensional walks. The types \textit{a}--\textit{e} are as explained in the text.
Each square gives the enumeration of walks with one dimension according to the row and the other according to column.
(*The three starred sequences enumerate walks of even length only,
returning to the point of origin.)}\label{tab:sum}
\end{table}

Walks of type \textit{aa} stay in one quadrant and return to the origin.
They are counted by \seqnum{A005568}~\cite[\Sect 4]{Guy},
\begin{align*}
\sum_{i} C_i C_{\frac n 2 -i} {n\choose 2i} &= C_{\frac n 2} C_{\frac n 2 +1} ,
\end{align*}
again for even $n$.

Walks of type \textit{ac}
stay in one quadrant but return to the abscissa (the promenade) and
are counted by \seqnum{A001700},
\begin{align*}
\sum_{i} C_i A_{n-2i} {n\choose 2i} &= {{2n+1}\choose n} .
\end{align*}
They
are discussed at length in \cite[\Sect 4]{Guy}.

Walks of type \textit{ce} are just restricted to the half-plane.
These are Guy's ``Sandsteps''~\cite{Guy}, introduced by Sands \cite{Sands},
and
are also counted by \seqnum{A001700}:
\begin{align*}
\sum_{i} C_i A_{n-2i} {n\choose 2i} &= {{2n+1}\choose n} .
\end{align*}

These and the remaining two-dimensional cases are summarized in Table~\ref{tab:sum}.
Most of these were investigated in \cite{DR,CMS,GKS}.
Their asymptotics were derived in~\cite{walks}.
More complicated walks involving diagonal steps have also been considered in the literature
(e.g.\@ \cite{Kreweras,Bousquet-Melou}).

\section{Restriction: Three dimensions}

Suppose that the swaggering pedestrian (or an intoxicated bird) can also move up (\U) or down (\D) at any point (and continue moving on those levels),
never venturing underground.
Suppose further that the path taken need only end up on the ground, not necessarily on the promenade.
The number of $n$-step walks of this type (\textit{ace}) is, by the general formula of the previous section,
\begin{align*}
\sum_{i,j}\frac{2^{n-2i-j}}{i+1}{2i\choose i}  {j\choose \div j 2}   {n\choose 2i,j,n-2i-j} &=
\sum_{i,j}\frac{2^{n-2i-j}}{i+1}    {n\choose i,i, \div j 2,\lceil j/2 \rceil,n-2i-j} .
\end{align*}

Table~\ref{tab:3D} provides computed initial terms for all three-dimensional walks with steps in all six directions (\N/\S/\E/\W/\U/\D) and with some requirement or other to return towards the origin.

\begin{table}
\centering\setlength\tabcolsep{3pt}
\begin{tabular}{|c|c|c|l|}
\bf Type & \bf Space & \bf Back & \head{Sequence} \\\hline\hline
\textit{aaa} & octant & origin & \footnotesize 1, 0, 3, 0, 24, 0, 285, 0, 4242, 0, 73206, 0, 1403028, 0, 29082339, \dots\@ (\seqnum{A064037}*) \\\hline
\textit{aab} & quad. & origin & \footnotesize 1, 0, 4, 0, 40, 0, 570, 0, 9898, 0, 195216, 0, 4209084, 0, 96941130, \dots \\\hline
\textit{aac} & octant & axis & \footnotesize 1, 1, 4, 9, 40, 120, 570, 1995, 9898, 38178, 195216, 805266, 4209084, \dots  \\\hline
\textit{aad} & octant & axis & \footnotesize 1, 1, 3, 7, 23, 71, 251, 883, 3305, 12505, 48895, 193755, 783355, 3205931, \dots \\\hline
\textit{aae} & quad. & axis & \footnotesize 1, 2, 6, 20, 74, 292, 1214, 5252, 23468, 107672, 505048, 2413776,  \dots\@ (\seqnum{A145867}) \\\hline
\textit{abb} & half & origin & \footnotesize 1, 0, 5, 0, 62, 0, 1065, 0, 21714, 0, 492366, 0, 12004740, 0, 308559537, \dots  \\\hline
\textit{abc} & quad. & axis & \footnotesize 1, 1, 5, 12, 62, 200, 1065, 3990, 21714, 89082, 492366, 2147376, 12004740,  \dots \\\hline
\textit{abd} & quad. & axis & \footnotesize 1, 1, 4, 10, 39, 131, 521, 1989, 8149, 33205, 139870, 592120, 2552155,  \dots \\\hline
\textit{abe} & half & axis & \footnotesize 1, 2, 7, 26, 108, 472, 2159, 10194, 49396, 244328, 1229308, 6273896, \dots \\\hline
\textit{acc} & octant & plane & \footnotesize 1, 2, 7, 24, 98, 400, 1785, 7980, 37674, 178164, 874146, 4294752, 21667932, \dots \\\hline
\textit{acd} & octant & plane & \footnotesize 1, 2, 6, 19, 67, 246, 947, 3746, 15213, 62950, 264920, 1129965, \dots\@ (\seqnum{A145847}) \\\hline
\textit{ace} & quad. & plane & \footnotesize 1, 3, 11, 44, 188, 842, 3911, 18692, 91412, 455540, 2306028, 11829424, \dots \\\hline
\textit{add} & octant & plane & \footnotesize 1, 2, 5, 14, 42, 132, 429, 1430, 4862, 16796, 58786, 208012, \dots\@ (\seqnum{A000108}) \\\hline
\textit{ade} & quad. & plane & \footnotesize 1, 3, 10, 36, 137, 543, 2219, 9285, 39587, 171369, 751236, 3328218, \dots\@ (\seqnum{A002212}) \\\hline
\textit{aee} & half & plane & \footnotesize 1, 4, 17, 76, 354, 1704, 8421, 42508, 218318, 1137400, 5996938, \dots\@ (\seqnum{A005572}) \\\hline
\textit{bbb} & full  & origin & \footnotesize 1, 0, 6, 0, 90, 0, 1860, 0, 44730, 0, 1172556, 0, 32496156, \dots\@ (\seqnum{A002896}*) \\\hline
\textit{bbc} & half & axis & \footnotesize 1, 1, 6, 15, 90, 310, 1860, 7455, 44730, 195426, 1172556, \dots\@ ($|$\seqnum{A138547}$|$) \\\hline
\textit{bbd} &  half & axis & \footnotesize 1, 1, 5, 13, 61, 221, 1001, 4145, 18733, 82381, 375745, 1703945, 7858225,  \dots \\\hline
\textit{bbe} & full  & axis & \footnotesize 1, 2, 8, 32, 148, 712, 3584, 18496, 97444, 521096, 2820448,  \dots\@ (\seqnum{A202814}) \\\hline
\textit{bcc} & quad. & plane & \footnotesize 1, 2, 8, 30, 138, 620, 3060, 14910, 76650, 390852, 2063376, 10832052,  \dots\\\hline 
\textit{bcd} & quad. & plane & \footnotesize 1, 2, 7, 25, 101, 416, 1787, 7792, 34645, 155722, 707795, 3242515, \dots\@ (\seqnum{A150500}) \\\hline
\textit{bce} & half & plane & \footnotesize 1, 3, 12, 53, 252, 1252, 6416, 33609, 178996, 965660, 5263728, 28936404 \dots \\\hline
\textit{bdd} & quad.  & plane & \footnotesize 1, 3, 11, 45, 195, 873, 3989, 18483, 86515, 408105, 1936881,  \dots\@ (\seqnum{A000984})  \\\hline
\textit{bde} & half  & plane & \footnotesize 1, 2, 6, 20, 70, 252, 924, 3432, 12870, 48620, 184756, 705432,  \dots\@ (\seqnum{A026375})  \\\hline
\textit{bee} &  full & plane & \footnotesize 1, 4, 18, 88, 454, 2424, 13236, 73392, 411462, 2325976,  \dots\@ (\seqnum{A081671})  \\\hline
\end{tabular}
\caption{Three-dimensional walks, 
required to return to the origin (three dimensions of type \textit{a} or \textit{b}), axis of origin (two), or plane of origin (one).
They may be constrained to a fraction of the space---octant (three of \textit{a}, \textit{c}, or \textit{d}), quadrant (two), or half-space (one), or else allowed the full space (zero). 
The types are as explained in the text.
(*The four starred sequences enumerate walks of even length only.
In one case, \textit{bbc}, the cited sequence, \seqnum{A138547}, has alternating signs.)}\label{tab:3D}
\end{table}

\subsection*{Acknowledgement}
I thank Jeffrey Shallit and a referee for helpful suggestions.

\bigskip
\hrule
\bigskip

\noindent
2010 \textit{Mathematics Subject Classification}: Primary 05A15.

\noindent
\textit{Keywords:} path enumerations, drunken walks, random walks, Dyck paths, lattice paths, Touchard's identity, Catalan numbers, Motzkin numbers, central binomial factors. 

\bigskip
\hrule
\bigskip

\noindent
(Concerned with sequences 
\seqnum{A000079},
\seqnum{A000108}, 
\seqnum{A000244},
\seqnum{A000302}, 
\seqnum{A000891}, 
\seqnum{A000984}, 
\seqnum{A001006}, 
\seqnum{A001405}, 
\seqnum{A001700}, 
\seqnum{A002212}, 
\seqnum{A002426}, 
\seqnum{A002894}, 
\seqnum{A002896},
\seqnum{A005566}, 
\seqnum{A005568}, 
\seqnum{A005572},
\seqnum{A005773}, 
\seqnum{A018224},
\seqnum{A026375},
\seqnum{A064037},
\seqnum{A081671},
\seqnum{A138547},
\seqnum{A145847},
\seqnum{A145867},
\seqnum{A150500},
\seqnum{A202814}.)

\noindent\hrulefill\medskip

\end{document}